\documentclass[12pt,preprint]{aastex}
\usepackage{color}
\shorttitle{Origin of Main Belt Comets}
\shortauthors{Haghighipour}

\begin{document}

\title{Dynamical Constraints on the Origin of Main Belt Comets}

\author{N. Haghighipour}
\affil{Institute for Astronomy and NASA Astrobiology Institute,\\
University of Hawaii-Manoa, Honolulu, HI 96822}
\email{nader@ifa.hawaii.edu}

\begin{abstract}

In an effort to understand the origin of the Main Belt Comets (MBCs) 
7968 Elst-Pizzaro, 118401, and  P/2005 U1, 
the dynamics of these three icy asteroids and a large number of
hypothetical MBCs were studied. Results of extensive  
numerical integrations of these objects suggest that these MBCs were formed in-place 
through the collisional break up of a larger precursor body. Simulations 
point specifically to the Themis family of asteroids as the origin of these objects
and rule out the possibility of a cometary origin (i.e. inward scattering of comets
from outer solar system and their primordial capture in the asteroid belt).
Results also indicate that while 7968 Elst-Pizzaro and 118401 maintain their orbits 
for 1 Gyr, P/2005 U1 diffuses chaotically 
in eccentricity and becomes unstable in $\sim 20$ Myr. The latter suggest that this MBC 
used to be a member of the Themis family and is now escaping away.
Numerical integrations of the orbits of hypothetical MBCs in the vicinity of the Themis
family show a clustering of stable orbits (with eccentricities smaller than 0.2 and
inclinations less than $25^\circ$) suggesting that
many more MBCs may exist in the vicinity of this family (although they might have
not been activated yet). The details of the results of 
simulations and the constraints on the models of the formation 
and origins of MBCs are presented, and their implications for 
the detection of more of these objects are discussed.
\end{abstract}

\keywords{minor planets, asteroids, methods: N-body simulations}

\section{Introduction}

The detection of comet-like activities in three icy asteroids, 
7968 Elst-Pizzaro, 118401, and P/2005 U1 (also known by their 
cometary indicators as 133P/Elst-Pizzaro, 1999 ${{\rm RE}_{70}}$, 
and Read) by \citet{Hsieh06} suggests the possibility of the 
existence of a new class of objects in the asteroid belt. 
Dubbed as Main Belt Comets, these objects have physical 
characteristics (e.g. comae and dusty tails) similar to those
of comets, whereas dynamically, they resemble asteroids (their
Tisserand numbers\footnote{For a small 
object, such as an asteroid, that is subject to the gravitational 
attraction of a central star and the perturbation of a planetary 
body P, the quantity 
${a_{\rm P}}/a \, + \,2 {[(1-e^2)\,a/{a_{\rm P}}]^{1/2}}\, \cos i$ 
is defined as its Tisserand number. In this formula, $a$ is the
semimajor axis of the object with respect to the star, $e$ is its 
orbital eccentricity, $i$ is its orbital inclination, and $a_{\rm P}$ 
is the semimajor axis of the planet. In the solar system, the 
Tisserand number of a small body with respect to Jupiter can be used 
to determine the cometary or asteroidal nature of its orbit. In general, 
the Tisserand numbers of comets with respect to Jupiter are smaller 
than 3, whereas those of asteroids are mostly larger.} 
with respect to Jupiter are larger than 3.) As indicated 
by their orbital elements (Table 1), these objects are close to 
the outer region of the asteroid belt and within or in the 
proximity of the Themis family of asteroids (Fig.1). The dusty 
tails of these bodies persist for many weeks during their perihelion
passages implying that tails of MBCs are not the ejections of 
impact-generated dust particles. An analysis of the dust trail of 
7968 Elst-Pizzaro by \citet{Hsieh04} suggests that small grains 
in the tail of this asteroid are dust particles that have been ejected from 
the surface of this body by the drag force of the gas produced
by the sublimation of near-surface water ice.

The dual characteristic of 7968 Elst-Pizzaro, 118401, and P/2005 U1
(i.e. their apparent cometary activities and Tisserand numbers larger than 3)
has raised many questions regarding the origin of these objects. On one hand, the 
physical appearance of these bodies may be taken as an evidence to argue that
MBCs are comets that were scattered inward from the outer regions of the solar system 
and were captured in orbits in the asteroid belt during the
early stages of the dynamical evolution of the solar system \citep{Levison09}.
On the other hand, the asteroidal-like Tisserand numbers of these bodies,
combined with their proximity to the Themis family of asteroids can be used to argue 
the in-place formation of these objects in or around their current orbits. This paper 
examines these scenarios by studying the dynamics of these three MBCs and the 
constraints that the results may apply to the origin of these objects. 

The currently known MBCs are a few kilometer in size. As indicated by 
Hsieh et al. (2004) and \citet{Hsieh06}, the rate of the reduction of the
sizes of these objects due to their cometary activities is approximately 
1 meter per year. Similar to comets, the activities
of MBCs are episodic and only intensify at their perihelion distances.
As a result, such a rate of size-reduction implies that MBCs may not have long
lifetimes and the current MBCs could not have started their
activities too long ago. It also suggests that many icy asteroids
might have had cometary activities in the past and are now inactive, 
and many more MBCs may exist in the outer region of the asteroid belt, which 
may be near their aphelia and have not started their activities yet.
This paper discusses these issues by 
studying the dynamics of a large number of hypothetical 
MBCs for different values of their orbital parameters, and by identifying the 
regions of the parameter-space where these objects have higher probability
of existence.

The outline of this paper is as follows. In section 2, the results of the 
numerical study of the dynamical properties of the 
three MBCs 7968 Elst-Pizzaro, 118401, and P/2005 U1
are presented. Section 3 has to do with 
the application of the results to the origin of these objects and their formation 
scenarios. Section 4 concludes this study by summarizing 
the results and discussing their implications for detections of more MBCs.

\section{Numerical Simulations}

As shown by \citet{Nesvorny98}, the asteroid belt, including
the locations of the currently known MBCs, is populated by many two-body 
mean-motion resonances (MMRs) with Jupiter. Fig.1 shows 
some of these resonances 
in the region around 7968 Elst-Pizzaro, 118401, 
and P/2005 U1. 
To explore whether these resonances affect
the long-term stability of these objects, the orbits of these bodies
were integrated for 1 Gyr. Integrations included all the planets
and Pluto, and treated MBCs as test particles. 
The latter is a realistic 
assumption based on the fact that the currently known 
MBCs are km-sized bodies \citep{Hsieh06} and at the current dynamical 
state of the asteroid belt, the probability of close encounters between 
objects of this size is negligibly small. The effects of non-gravitational 
forces such as Yarkovsky, and the effect of the mass-loss of MBCs 
due to their cometary activities were not included. 
These effects will be discussed in more detail in a future article.
Integrations were carried out with Bulirsch-Stoer 
and with the Second-Order Mixed-Variable Symplectic (MVS)
integrators in the N-body integration package MERCURY \citep{Chambers99}.
Bulirsch-Stoer is a non-symplectic general integrator,
which unlike MVS, has a self-adjusting variable timestep to maintain
accuracy. The reason for using two different integrators is to assure that the results were
independent of the choice of the integrator.
The initial orbital elements of the MBCs and those of the 
planets were obtained from documentation on solar system dynamics 
published by the Jet Propulsion Laboratory 
(http://ssd.jpl.nasa.gov/?bodies). 
The timestep of each integration was set to 9 days.\footnote{Note that for
the Bulirsch-Stoer integrator, this would be the stepsize used
for the first timestep only.}

Fig.2 shows the results of the simulations. As shown by the left
panel of this figure, 7968 Elst-Pizzaro and 118401 maintain their orbits 
for 1 Gyr. However, P/2005 U1 becomes unstable after approximately 20 Myr. 
To obtain a more reliable estimate of the lifetimes of these 
objects, integrations were also carried out for different initial values of the 
semimajor axes and eccentricities of these bodies by changing these quantities
in increments of $\Delta a=0.0001$ AU and $\Delta e= 0.001$
within the ranges of their observational uncertainties.
Results confirmed the long-term stability of 7968 Elst-Pizzaro and 118401, whereas
P/2005 U1 became unstable in all simulations with a median lifetime of $\sim$57 Myr. 

The graphs on the right panel of Fig.2 show the variations of the semimajor
axes and eccentricities of 7968 Elst-Pizzaro, 118401, 
and P/2005 U1 during the time of integration. As shown here,
the semimajor axes and eccentricities of 7968 Elst-Pizzaro and 118401
stay within a small region outside the influence zone of the 2:1 MMR 
with Jupiter, whereas those of P/2005 U1 vary in the close proximity of
this region. It is necessary to emphasize that the penetration of P/2005 U1
into the region of 2:1 MMR, as shown in Fig.2, is only apparent. This object
does not get captured in the 2:1 MMR with Jupiter. The boundary of the 2:1
resonance in this figure has been shown merely for the purpose of portraying
the occasional proximity of P/2005 U1 to this region.
The latter implies that similar to its current state where this object
is the closest of the three MBCs to the boundary of the 2:1 MMR, 
P/2005 U1 was perhaps {\it originally} the closest MBC to this resonance as well.
The original proximity of P/2005 U1 to the 2:1 MMR has
resulted in a gradual increase in its orbital eccentricity which will eventually cause 
its orbit to become unstable.

As mentioned in the Introduction, MBCs are only likely to be active for a short time.
Long-term integrations of the motions of these objects, as the one 
mentioned above, may extend the calculations to beyond the duration 
of the activation of these bodies. However, given that (1) many MBCs,
which have not started their activations yet, may exist in the vicinity 
of the currently known ones, and (2) many MBCs that existed in that region
in the past might have been scattered to the other parts of the solar system 
where they could become active and detected, it proves useful to carry out 
long-term integrations of the orbits of these bodies and explore the dynamical properties
of their orbital space. For this reason, 
the orbits of a large number of hypothetical MBCs were integrated for 100 Myr  
for different values of their semimajor axes, eccentricities and inclinations.
The initial semimajor axes of these bodies were varied 
systematically from 3.14 AU to 3.24 AU in increments of 0.01 AU. Their 
initial eccentricities were taken to be between 0 and 0.4 with $\Delta e=0.01$, and
their initial orbital inclinations were chosen from a range of 0 to $40^\circ$
with $\Delta i = 0.5^\circ$. Other angular variables of these objects (i.e. mean anomaly,
argument of perihelion, and longitude of ascending node) were chosen to be
zero. Fig. 3 shows the initial distribution of these bodies
(the combined purple and green circles) and the final results. 
The green circles represent the initial conditions 
that correspond to stable orbits for the duration of the integrations, whereas 
purple indicates instability. To prevent overcrowding of the graphs, only
objects corresponding to $\Delta e =0.02$ and
$\Delta i = 2.5^\circ$ have been shown. The separation of the stable and unstable regions
will not change if all objects are included. As shown in this figure, 
the semimajor axes and eccentricities of 7968 Elst-Pizzaro 
and 118401 place them in the stable region whereas those
of P/2005 U1 are close to the unstable part of the graph. Fig 3. also shows that 
for a given value 
of the semimajor axis of an MBC, the state of stability depends on 
the values of its initial eccentricity and orbital inclination. 
As expected, similar to main belt asteroids, objects
with initial inclinations larger than $\sim 25^\circ$ 
became unstable due to interactions with the $\nu_5$, $\nu_6$ and
$\nu_{16}$ secular resonances, and the Kozai resonance. For lower inclinations, orbital 
stability is driven by the values of the apastron distances of these
objects. Those hypothetical MBCs close to or inside 
the 2:1 MMR with Jupiter became 
unstable in a short time. An analysis of the orbits of the unstable objects 
indicates that approximately $80\%$ of these bodies were scattered to 
large distances outside the solar system. This is a familiar result that 
has also been reported by \citet{OBrien07} and \citet{Hagh08} in their
simulations of the dynamical evolution of planetesimals in the outer 
asteroid belt. From the remaining $20\%$ unstable MBCs, approximately $15\%$ 
collided with Mars, Jupiter, or Saturn, and a small fraction $(\sim 5\%)$ 
reached the region of 1 AU.

\section{Implications for the Origin of MBCs}

As mentioned earlier, the cometary appearance of 7968 Elst-Pizzaro, 
118401, and P/2005 U1,  and the values of their Tisserand numbers
point to two scenarios for the formation and
origin of these objects. While the latter favors in-situ formation 
(as members of the Themis family),
the former implies the possibility of the inward scattering and primordial capture of cometary 
bodies from the reservoirs of icy materials at the outer region of the 
solar system.\footnote{A third possibility, i.e, the 
MBCs are recent comets from the Kuiper belt or Oort cloud
and have been captured in orbits in the main belt, can be ruled
out since as shown by \citet{Fernandez02}, numerical simulations
have not been able to reproduce the transfer of comets from those
regions to the main belt at the current dynamical state of the solar system.} 
In the following, the plausibility of each of these two scenarios is discussed.

It has recently been suggested that the scattering of
cometary objects into orbits in the asteroid belt could 
have been possible at the early stages of the dynamical evolution of 
the solar system. 
Numerical simulations by \citet{Gomes05}, \citet{Morbidelli05}, and 
\citet{Tsiganis05} have indicated that the migrations of giant planets, 
and the subsequent passage of Jupiter and Saturn through the 
2:1 mean-motion resonance in an early epoch, could have affected the 
motions of many comets beyond the orbit of Neptune and scattered them 
into orbits in the outer region of the asteroid belt. 
Recent simulations by \citet{Levison09} show that,
within the context of this model, many trans-Neptunian planetesimals
might have been scattered inwards and captured in the regions of 
Trojans and Hilda asteroids, as well as in orbits as close in as  2.68 AU.
Despite the instability of many of these bodies (majority of the scattered 
objects have large eccentricities and inclinations), a small fraction 
of these planetesimals maintain low eccentricity and/or low 
inclination stable orbits. 

Whether these bodies can be the sources of MBCs is, however, uncertain. 
The results of the simulations by \citet{Levison09} show a high efficiency
for the delivery of D-type and P-type asteroids, in particular to the region of Trojans.
However, because C-type asteroids are mostly concentrated towards smaller
semimajor axes, and also because it is not known how different
the orbital distribution of C-type and D-type asteroids are, the results by \citet{Levison09} 
may not be able to give information about the capture of C-type asteroids.
Additionally, if MBCs were objects scattered inward from Kuiper belt,
they would be expected to be optically red, whereas
observations by \citet{Hsieh06}, and \citet{Hsieh08} have indicated that MBCs
are C-type asteroids with no specific optical color. 

The alternative scenario, that is, MBCs are bona-fide asteroids
that were formed through break up of their progenitor,
is, however, consistent with the orbital properties and the 
spectral classification of these objects.
In this scenario, the break up of the precursor asteroid could have produced 
many km-sized fragments, among which only those with low
inclinations and low eccentricities maintained their orbits for
long times. The orbits of these stable fragments are naturally
asteroidal (i.e. their Tisserand numbers are larger than 3),
and as indicated by the simulations of section 3,
many of these bodies could currently exist on stable orbits
in the outer region of the asteroid belt. In regard to  7968 Elst-Pizzaro, 118401, 
and P/2005 U1, this scenario points to the Themis family as the origin of
these objects.  The orbital proximity of these bodies to one another
and to the Themis family of asteroids, and the fact that they are spectrally C-type,
are also consistent with this scenario. In fact, recent simulations of the dynamics of the 
members of the Themis family have indicated the presence of a  
smaller $\sim 10$ Myr sub-family (known as Beagle) with 27 asteroids 
within these objects \citep{Nesvorny08}. 
The bottom graph of Fig.3 shows the members of this family and the locations of
7968 Elst-Pizzaro, 118401, and P/2005 U1. As shown in this figure, 
7968 Elst-Pizzaro could potentially
be one of the members of the Beagle family.

It is important to emphasize that the orbital clustering of MBCs, 
as an evidence for connecting their origins to the Themis family of asteroids,
needs to be taken with some caution. The proximity
of 7968 Elst-Pizzaro and 118401, and their orbital similarities,
are not surprising. The latter was detected through an observational 
survey that specifically targeted asteroids in the vicinity of 
7968 Elst-Pizzaro \citep{Hsieh06}. The MBC P/2005 U1 was, however,
discovered in an un-targeted survey, and its proximity to the members
of the Themis family was unexpected. This implies that MBCs might also exist
elsewhere in the asteroid belt. In fact, the recently discovered 
fourth MBC, P/2008 R1 (Garradd), with a semimajor axis of 2.726 AU 
\citep{Jewitt09}, is one of such objects. However, the short lifetime of 
this MBC, and that its immediate surrounding in the $(a-e)$ space
is unstable, suggests that this object might have formed in another 
location in the asteroid belt, and reached its current
orbit through interactions with giant planets.

Although the collision/break up of large asteroids as a mechanism 
for the formation of MBCs portrays a 
consistent picture of the dynamical properties of these objects,
it does not naturally account for the comet-like 
activities of these bodies (i.e. the comet-like tails of these bodies are not the 
natural results of such impacts).
In an analysis of the dust trail of 7968 Elst-Pizzaro, \citet{Hsieh04}
suggested that the dusty coma and tail of this MBC
have formed through the interactions of dust grains on the
surface of this body with the gas produced by the sublimation
of near-surface water ice. This suggestion is based on the assumption
that, 7968 Elst-Pizzaro, despite its low orbital eccentricity, which
gives it a fairly constant solar heating rate,
has been able to retain some fraction of its original water ice, 
in particular in small depths beneath its surface.  
Note that similar to the
members of the Themis family, 7968 Elst-Pizzaro, 118401, and P/2005 U1
are C-type asteroids and are volatile-rich. 
Recent calculations by \citet{Schorghofer08} have shown that 
asteroids in the region between 2 AU and 3.3 AU can 
retain their water ice for as long as the age of the solar system
if their surfaces are covered by a layer of dust, even as thick as 
only a few meters. Given that the MBCs 7968 Elst-Pizzaro, 118401, 
and P/2005 U1 are in
close proximity of one another, the analyses by Hsieh et al. (2004) 
and \citet{Schorghofer08} are applicable to all three of them and
suggest that the cometary tails of these objects could
be due to the sublimation of sub-surface water ice that has been 
exposed through collisions of these bodies with small meter-sized objects 
\citep{Hsieh06}.  It also implies that many more MBCs may exist in the outer part of the 
asteroid belt, which either have collided with meter-sized bodies and are
on their ways to their perihelia where they start
their comet-like activities, or they are still awaiting to undergo
activation-triggering collisions.

\section {Concluding Remarks}  

The orbits of the three icy asteroids, 7968 Elst-Pizzaro, 118401, 
and P/2005 U1, and a large number of hypothetical MBCs were 
integrated numerically for different values of their orbital 
elements. Results indicated that 7968 Elst-Pizzaro and 118401 
are stable for 1 Gyr whereas P/2005 U1 becomes unstable in
approximately 20 Myr. The chaotic diffusion of this object in eccentricity,
which is the result of its current orbital proximity to the region of 2:1 MMR with Jupiter,
suggests that this object was originally formed close to the 2:1 resonance.

The results of the dynamical simulations presented in this study apply
certain constraints to possible scenarios for the formation and origin
of MBCs. The proximity of the orbits of these objects
to one another and to the members of the Themis and Beagle families
of asteroids, combined with their spectral type, favors the idea that 7968 Elst-Pizzaro, 118401, 
and P/2005 U1 were formed in-place as members of the Themis family 
(i.e. through the collisional break up of Themis parent body). 
Also, simulations show that the orbits of stable hypothetical MBCs cluster around the Themis
family implying that many more MBCs may presently exist in that region.
The alternative scenario, that is, the inward scattering of cometary 
objects from distances outside the solar system and their primordial capture
in asteroidal orbits, although capable of delivering objects into
stable orbits, may not be able to portray a clear picture of the capture of C-type asteroids
during the early dynamical evolution of the solar system.
This scenario is also inconsistent with the optical
colors and spectral features of MBCs. Unlike the comets from the 
Kuiper belt, which are optically red, MBCs are C-type asteroids with
no specific optical color \citep{Hsieh06,Hsieh08}. 
The latter also points to the implausibility of the assumption that
the parent body of the Themis family itself might have been a
captured comet.

Simulations of the dynamics of hypothetical MBCs indicated that
as expected, objects close to the edge of 2:1 MMR 
with Jupiter became unstable with lifetimes ranging from a few hundred
thousand years to more than 90 Myr. For a given set of initial values of $a, i$, and
angular variables, the median lifetime of these objects became shorter by
increasing the initial values of their orbital eccentricities. For the ranges of the 
values of the semimajor axis and inclination considered here (i.e. 3.14 - 3.24 AU, and
0 - 0.4, respectively), this median
lifetime ranged between 20 to 50 Myr. It is necessary to emphasize that
in general, stability for 100 Myr may not be indicative of a stable orbit for the
age of the solar system. However, given the small km-sized sizes of MBCs,
and their rates of mass-loss, stability for 100 Myr will be sufficient to identify
the boundaries of the stable region.

The effects of non-gravitational forces such as Yarkovsky, and the
effect of the mass-loss of MBCs due to their cometary activities 
were not considered in this study. Given that the activation of MBCs 
is mainly limited to the duration of their perihelion passages, which is
short compared to the orbital periods of these objects, the effects of 
these perturbations do not seem to disturb the dynamical picture of
7968 Elst-Pizzaro, 118401, and P/2005 U1 
as portrayed in this study. The results of detailed simulations of these 
effects will be published in another article.

A dynamical property of 7968 Elst-Pizzaro, 118401, and P/2005 U1 that 
supports their in-place formation, as the more probable scenario 
for the origin of these objects, was the orbital proximity of these bodies 
to one another and to the Themis family of asteroids. 
Although the proximity of 7968 Elst-Pizzaro and 118401 was the
result of a targeted observational survey, in regard to the 
detection of more MBCs, this suggests that the more likely places to 
observe these objects may be families of asteroids with large
parent bodies capable of differentiating and forming ice-rich mantles, 
in particular within the outer half of the asteroid belt. 
It is necessary to note that, 
as indicated by the results of the dynamical simulations 
in section 3, some members of such families may 
interact with giant planets and reach 
orbits in other regions of the asteroid belt.
These objects, although originally from an MBC-forming family, may 
appear as stand-alone MBCs. The recently discovered MBC, 
P/2008 R1 (Garradd), may be one of such objects.
Telescopes such as Pan STARRS 1 with the 
capability of continuous and large-scale monitoring of the sky, 
would be capable of detecting such individual MBCs, and are ideal for 
carrying out targeted surveys for families of these objects.

\acknowledgements
I gratefully acknowledge fruitful discussions
with H. Hsieh, D. Jewitt, H. Levison, K. Meech, D. Nesvorny, 
and N. Schorghofer. I would also like to thank D. P. O'Brien
and another (anonymous) referee for reviewing the article and for their 
valuable suggestions that resulted in the improvement of this manuscript. 
This work was partially supported by the NASA 
Astrobiology Institute under Cooperative Agreement NNA04CC08A at the 
Institute for Astronomy, and by the office of the Chancellor of the 
University of Hawaii, and a Theodore Dunham J. grant administered 
by Funds for Astrophysics Research, Inc.

\clearpage
\begin{figure}
\vskip -3.5in
\plotone{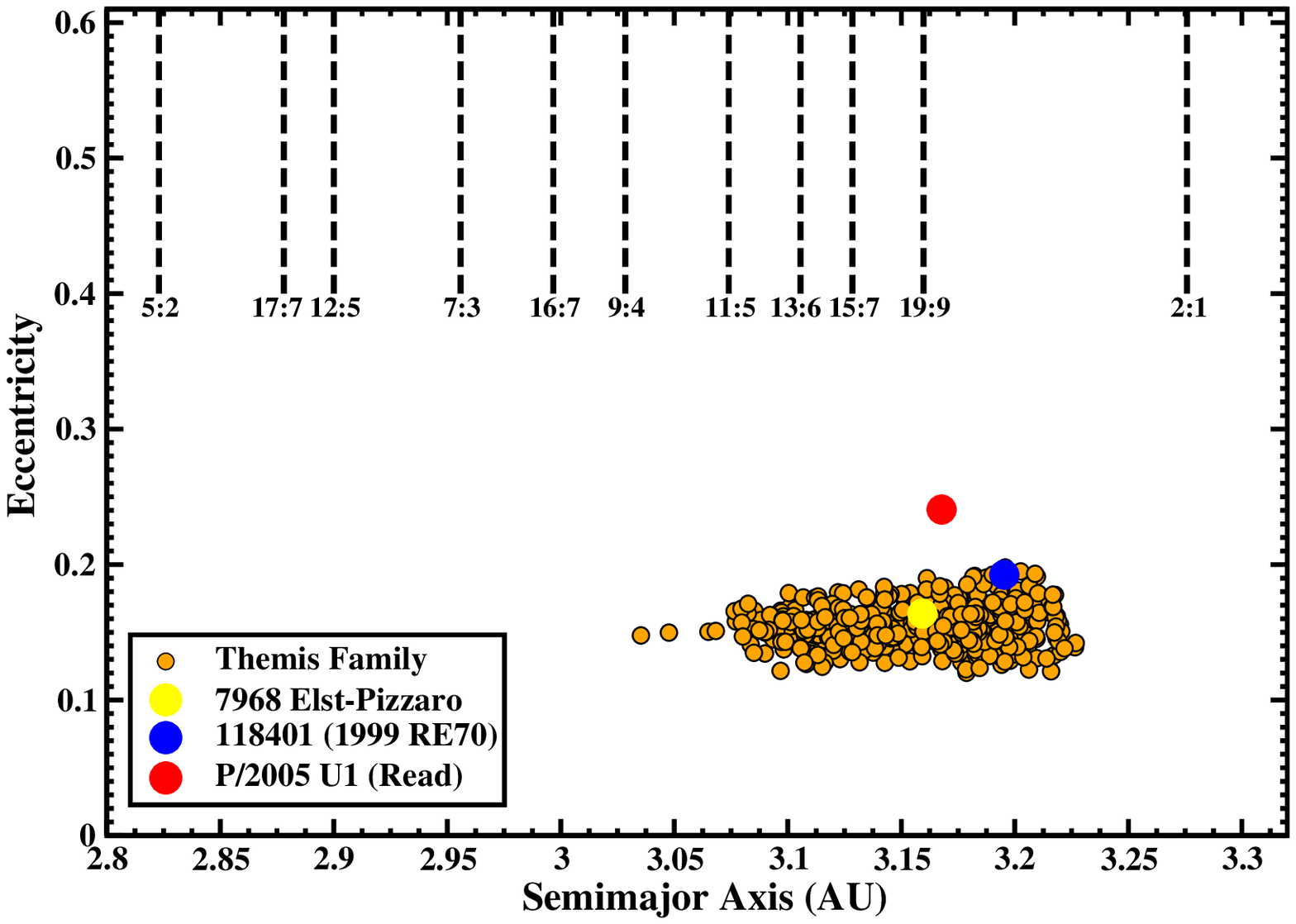}
\vskip -1.4in
\caption{The three MBCs 7968 Elst-Pizzaro, 118401, P/2005 U1, 
and the Themis family of asteroids in the $(a-e)$ plane. As shown here,
7968 Elst-Pizzaro and 118401 are within the Themis
family whereas P/2005 U1 is in its proximity. The value of $V_{\rm cutoff}$
(the parameter that is used to define the relative spread of the family in 
the $a, e, i$ space when using a family-finding search algorithm) of the
Themis family shown here is 100 m/s.
\label{fig1}}
\end{figure}

\clearpage
\begin{figure}
\plotone{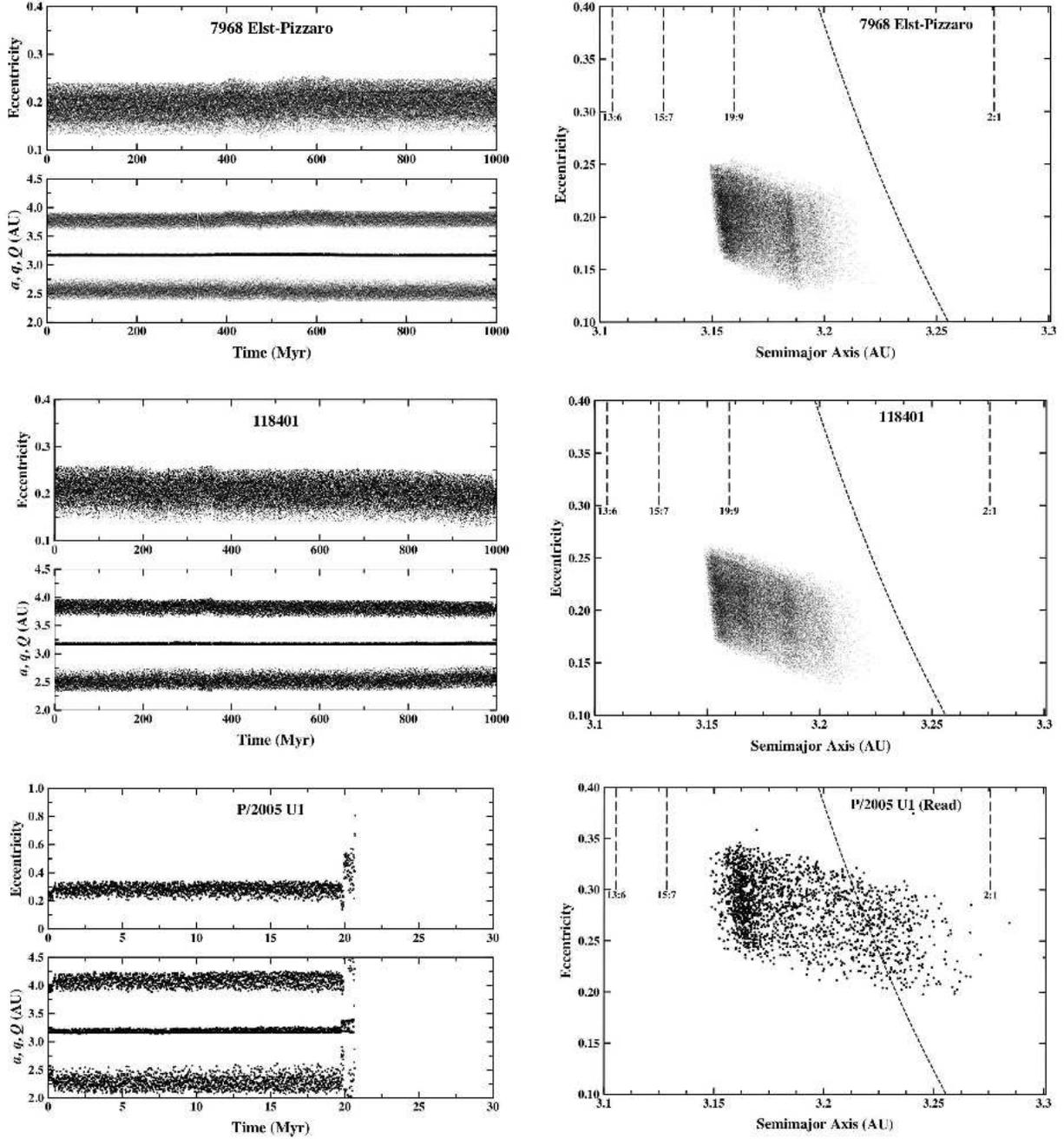}
\caption{Left: graphs of the eccentricities, semimajor axes $(a)$,
perihelion $(q)$, and aphelion $(Q)$ distances of 
7968 Elst-Pizzaro, 118401, and  P/2005 U1. As shown here,
7968 Elst-Pizzaro and 118401 are stable for 1 Gyr whereas P/2005 U1
becomes unstable after approximately 20 Myr.
Right: the region of the $(a-e)$ plane
occupied by each MBC during the time of integrations. The dashed
line in each graph shows the inner edge of the influence zone of the 2:1
MMR with Jupiter. As shown here, while the variations of $a$ and $e$ for
7968 Elst-Pizzaro and 118401 stay outside the resonant region,
the eccentricity and semimajor axis of  P/2005 U1 show an apparent
penetration into the 2:1 MMR causing the orbit of this object to become unstable.
\label{fig2}}
\end{figure}

\clearpage
\begin{figure}
\plotone{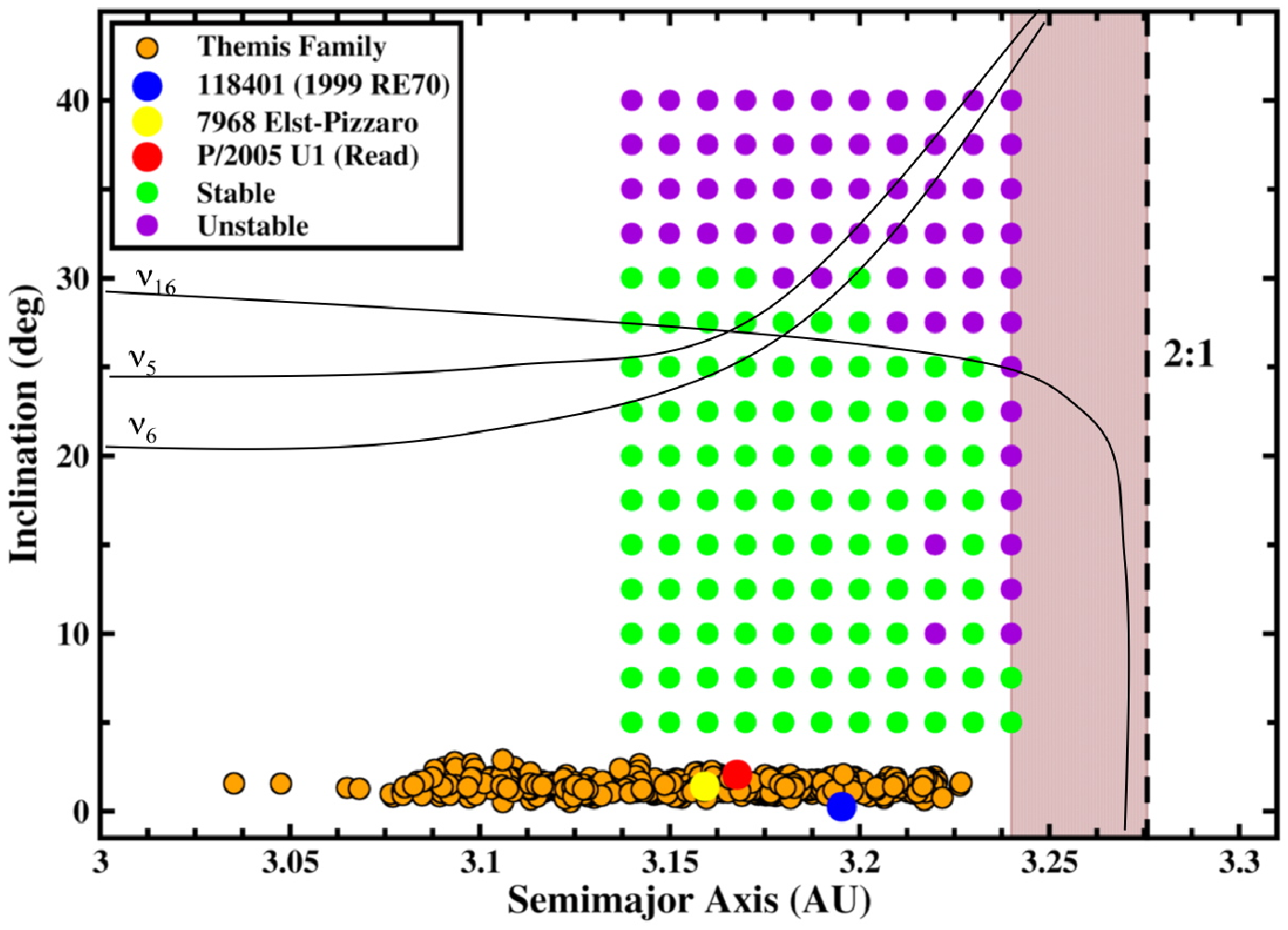}
\plotone{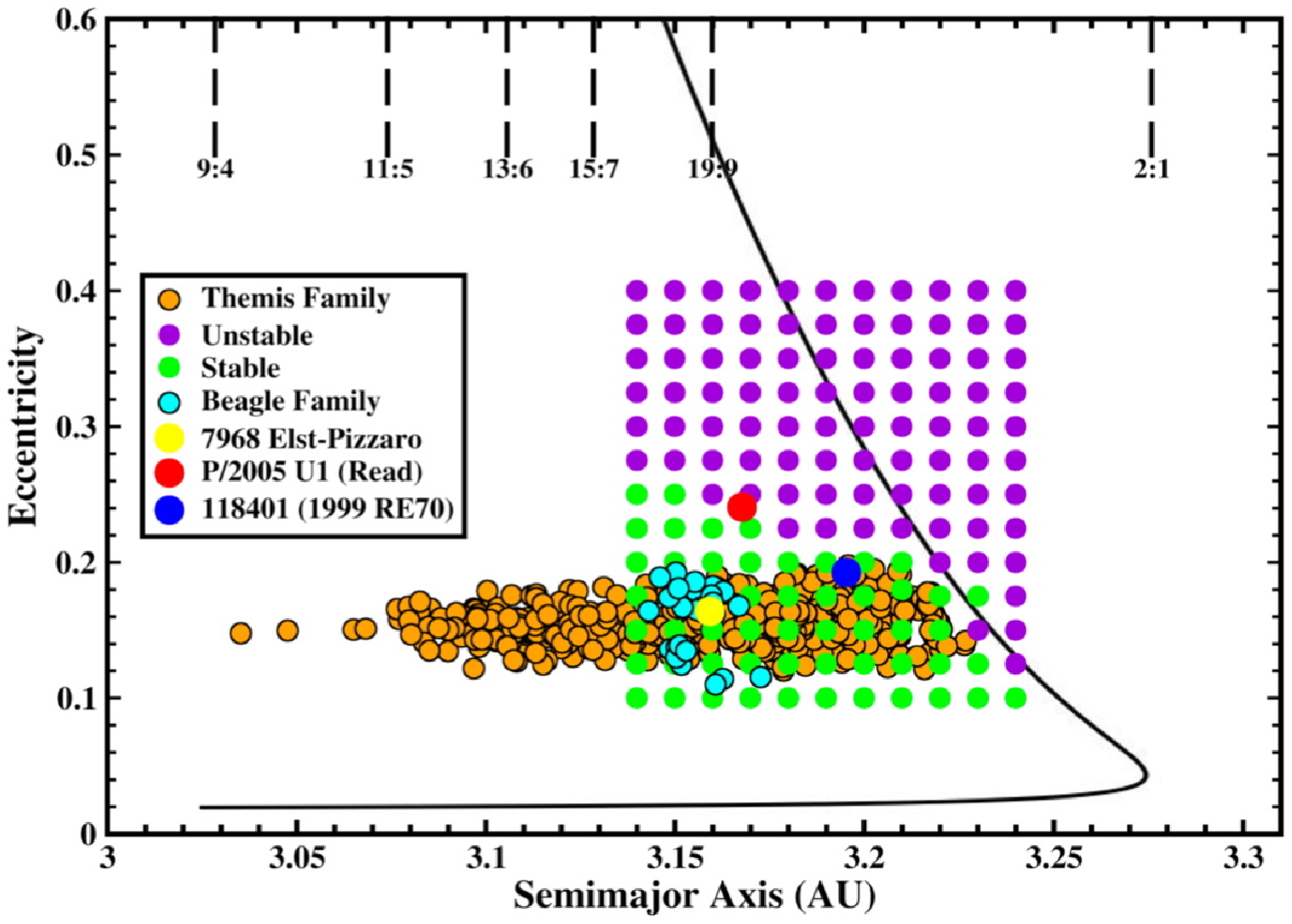}
\vskip -2.73in
\caption{Top: graph of the stability of hypothetical MBCs 
in terms of their inclinations. 
As an example, the regions of secular resonances $\nu_5$, $\nu_6$, and $\nu_{16}$
corresponding to an eccentricity of 0.1 are also shown.
Bottom: graph of the stability of hypothetical MBCs
in terms of their eccentricities. 
The brown area in the top graph and solid line in the bottom graph show 
the region of the 2:1 MMR with Jupiter. Each circle represents the
initial orbital elements of an object. Circles in green correspond
to stable ones whereas those in purple show instability. 
As shown here, similar to the asteroid in the main belt, objects with inclinations larger 
than $\sim25^\circ$ and eccentricities larger than $\sim 0.2$ are unstable.
\label{fig4}}
\end{figure}

\clearpage
\begin{deluxetable}{lccccc}
\tablewidth{0pt}
\tablecaption{Orbital Elements of MBCs
\citep{Hsieh06}}
\tablehead{
\colhead{MBC} &
\colhead{$a$ (AU)} &
\colhead{$e$} &
\colhead{$i$ (deg)} &
\colhead{Tisserand} &
\colhead{Diameter (km)}} 
\startdata
7968 Elst-Pizzaro     & 3.156  & 0.165  & 1.39  & 3.184  & 5.0 \\
118401  & 3.196  & 0.192  & 0.24  & 3.166  & 4.4 \\
P/2005 U1 & 3.165  & 0.253  & 1.27  & 3.153  & 2.2 \\

\enddata
\end{deluxetable}

\end{document}